\documentclass[prl,twocolumn,showpacs,preprintnumbers,amsmath,amssymb]{revtex4-1}

\usepackage{graphicx}
\usepackage{dcolumn}
\usepackage{bm}
\usepackage{textcomp}
\usepackage{wasysym}

\begin{document}



\title{Thermal Conductance of Ballistic Point Contacts}

\author{Th.\ Bartsch, M.\ Schmidt, Ch.\ Heyn, and W.\ Hansen}
\affiliation{Institut f\"ur Angewandte Physik und Zentrum f\"ur
Mikrostrukturforschung,\\Jungiusstra\ss e 11, D-20355 Hamburg,
Germany}

\begin{abstract}
We study the thermal conductance of ballistic point contacts. These contacts are realized as few nanometer long pillars in so called air-gap heterostructures (AGHs). The pillar length being much smaller that the mean free path of the phonons up to room temperature. Due to the small dimension and the low density of the pillars the thermal conductance of the AGHs is several orders of magnitude reduced in comparison to bulk structures. The measurement results are in quantitative agreement with a simple model that based on the Boltzmann transport equation.
\end{abstract}

\maketitle
Recent studies established significant influences on the thermal transport in solids by nanostructuring \cite{Sch00,Hoc07,Har02, Hsu04, Pou08, Ven01}. This opens the doorway for the realization of novel thermoelectric devices, where a low thermal conductance is essential \cite{Hoc07,Har02, Hsu04, Pou08, Ven01, Vin10, Mar04, Zeng06, Wes08}. In particular, using nanostructuring it is possible to scale the dimension of a solid to a regime where the mean free path (mfp) of the heat carrying phonons is larger than both the diameter and the length of the structure of interest \cite{Sch00}. The associated thermal transport regime is named ballistic. So far there are many theoretical studies of ballistic thermal-transport \cite{Ang98, Che98, Pra06, Pra07, Zhou09}, which  all emanate from the Boltzmann transport equation,  but there are only few experimental data. Schwab et al. described experiments on suspended microstructures \cite{Sch00}, but in those experiments the phonon mfp is only at very low temperatures lager than the dimension of the structure. Furthermore, a kind of ballistic thermal-transport was observed in superlattices \cite{Che98}, since in the cross-plane direction the phonon mfp is longer than the periodicity of the superlattice. However, in the in-plane direction of the superlattices still phonon scattering takes place.  In Ref. \cite{Hey10} we have demonstrated the fabrication of so called air-gap heterostructures (AGHs)  that enable the preparation of pillars with  controlled length of a few nanometers. In the following, we demonstrate that these pillars represent pure ballistic thermal point contacts in a wide temperature range up to room temperature.

\begin{figure}[b]
\centering
\includegraphics[width=7cm]{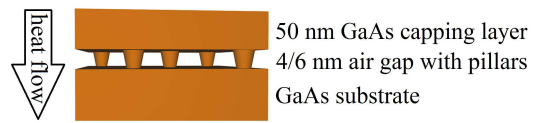}
\caption{Scheme of the investigated structure.}
\label{Fig1}
\end{figure}

The AGHs are composed of a 50\,nm GaAs capping layer that is supported  by a number of nanopillars over a GaAs substrate. Here we investigate nanopillars with length of 4\,nm and 6\,nm, respectively. A schematic drawing of an AGH is shown in Fig.\,\ref{Fig1}.  The structures were fabricated using molecular beam epitaxy (MBE) with a combination of in situ  local droplet etching (LDE) \cite{Hey094,  Hey09, Hey11,  Ste08, Hey092} and ex situ selective chemical wet-etching of a sacrificial layer.
 In the first MBE growth step an AlAs layer with thickness of 4\,nm or 6\,nm was deposited on a (001) GaAs substrate. This is followed by a Ga droplet etching  step inside the MBE growth chamber, which generates in a self-organized fashion nanoholes deeper than the thickness of the AlAs layer \cite{Hey094, Hey092, Hey11, Hey09, Ste08}. These holes were filled  by subsequent growth of a 50\,nm thick GaAs layer. After removal of the sample from the MBE setup, a mesa was fabricated as described below. Finally, the AlAs layer was selectively removed by etching with a 5$\%$ solution of hydrofluoric  acid, leaving  the filled holes as GaAs nanopillars. Because of the MBE fabrication, the  pillars are almost defect free and lattice matched to the substrate. We note that the capping layer keeps the perfect epitaxial relationship to the substrate crystal lattice. A more detailed description of the preparation technique and a verification that the capping layers are separated from the substrate by the pillars was given in Ref\,\cite{Hey10}.
 
The thermal conductance was measured using the 3$\omega$ method. This method has been developed for  measuring  the thermal conductance of bulk materials and thin films \cite{Cah89,  Cah94, Bor01}. A thin metal stripe on top of a specimen is used as heater and temperature sensor simultaneously. On our AGH samples  we prepared a heater stripe with 1.6\,mm length and  15\,$\mu$m width by optical lithography and evaporation deposition of 30\,nm Au. Afterwards, the semiconductor material aside from the heater was removed to a depth of 100 nm in a wet-etching step using the metal stripe as an etch mask. This prevents lateral heat spreading in the structure and opens the AlAs sacrificial layer for subsequent selective chemical wet-etching. 
 
For 3$\omega$ measurements, a sinusoidal ac current with frequency $\omega$ was driven through the heater. The generated temperature oscillations induce a resistance oscillation at frequency 2$\omega$. Thus the voltage drop across the heater includes a component at the frequency 3$\omega$. Measurements of this 3$\omega$ voltage enables the determination of the temperature rise $\Delta T$ and finally of the thermal conductance of the material below the heater \cite{Cah89, Cah94, Bor01}. In the samples studied here, the AGHs are treated as a thin film between heater and substrate. Cahill et al.  discussed such an experimental situation \cite{Cah94, Bor01}. There it is assumed that the magnitude of the temperature rise $\Delta T$ is the sum of contributions from the film ($\Delta T_{f}$) and the substrate ($\Delta T_{s}$), if the film represents a significant thermal resistance $R_{th}$ \cite{Cah94, Bor01}:
 \begin{eqnarray}
 \Delta T=\Delta T_{f}+\Delta T_{s}. 
 \label{Eq1}
\end{eqnarray}
Below it will be described how the contribution $\Delta T_f$ can be determined from the measured $\Delta T$. Once $\Delta T_f$ is known, the thermal conductance $K_f$ of the film can be
calculated from the heating power $P$ with the relation
\begin{eqnarray}
  \Delta T_{f}=R_{th}I_{th}=\frac{P}{K_f}.
 \label{Eq2}
\end{eqnarray}
that is given by the general equation for a heat current $I_{th}$ \cite{Cah94}. Note that the right side of the equation is only valid if the heat current $I_{th}$ is equal to the power $P$ impressed by the heater, i.e., heat losses are neglected \cite{Cah94}.

\begin{figure}[bt]
\centering
\includegraphics[width=8.5cm]{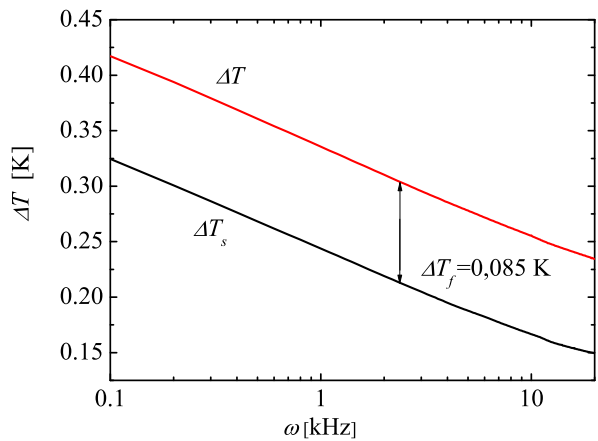}
\caption{Temperature rise as function of frequency determined from 3$\omega$ measurements at $T$=300\,K  on an AGH sample with 6\,nm long pillars. The black line represents the measurement before and the red line after selective removal of the AlAs layer.}
\label{Fig2}
\end{figure}

In previous publications, $\Delta T_{f}$ was determined with a differential technique  by measuring one sample with the investigated film ($\Delta T$) and a second reference sample with a heater directly on the substrate ($\Delta T_{s}$) \cite{Cah94, Bor01}. In our experimental approach we are able to execute both measurements on the same sample. The first measurement is performed before etching of the sacrificial AlAs layer and the second afterwards.  The temperature rise on the unetched sample is expected to be equal to the temperature rise $\Delta T_{s}$ on a GaAs substrate, since the specific thermal resistance of the thin AlAs is nearly identical compared to the substrate.  To verify this we performed measurements on the unetched sample and an intrinsic GaAs substrate, which exhibit no significant differences.  In the next step, after determining $\Delta T_{s}$ on the unetched sample, we selectively etched the AlAs  to uncover the pillars. Now the 3$\omega$ measurements were repeated to determine $\Delta T$. Figure\,\ref{Fig2} shows exemplarily, how the etching influences the temperature rise. 
As expected, $\Delta T$ is significantly larger after etching. We associate the difference $\Delta T_f$ to the higher thermal resistance of the AGHs after etching. The slopes of the temperature rises $\Delta T$ and $\Delta T_{s}$ are nearly equal. This is expected and indicates that the slopes are not affected by the presence of the film below the heater, i.e., the slope is just influenced by the substrate \cite{Cah94, Bor01}.  Using the relation between the slope and the thermal conductivity of the substrate established by Cahill et al. in Ref.\,\cite{Cah89}, we determined the thermal conductivity of the substrate. The results are in good agreement with the literature values for GaAs thermal conductivity \cite{Car65, May66}. This indicates  that the 3$\omega$\,method is applicable to the AGH samples studied here.  

We determined the thermal conductance of the pillar ensembles in the AGHs in the temperature range between 20\,K and 300\,K with eq.\,(\ref{Eq2}) from the measured 3$\omega$ data.  The spreading of our data at comparable conditions is less than 15\,$\%$, which is smaller than the error of 20\,$\%$ estimated by Borca-Tasciuc et al. \cite{Bor01} for 3$\omega$ measurements on thin films by using the differential technique with two samples.
\begin{figure}[tb]
\centering
\includegraphics[width=8.5cm]{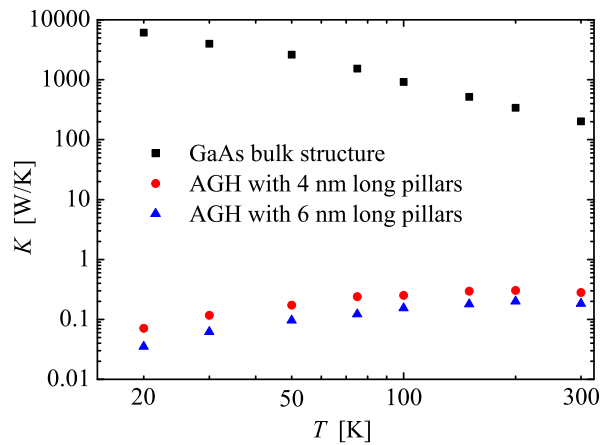}
\caption{Thermal conductance of two AGH samples and a GaAs bulk structure (black squares) calculated as reference as described in the text. The red points correspond to an AGH with 4\,nm long pillars and the blue triangle to an AGH with 6\,nm long pillars. The size of the symbols is equal to the error bars in the logarithmic plot.}
\label{Fig3}
\end{figure}
Figure\,\ref{Fig3} presents conductances of AGHs with 4\,nm and 6\,nm long pillars. Clearly visible is an increase of the thermal conductance of the AGHs with increasing temperature. For temperatures higher than 150\,K, the measured values saturate.  In average the thermal conductance of the 4\,nm AGH is 1.75 times higher than that of the 6\,nm AGH.  For comparison, we calculate the thermal conductance of GaAs bulk material with dimensions of the air gaps. Therefor we neglected any surface effects, i.e., we just multiplied the thermal conductivity of the substrate by the area of the heater and divided it by the average length of 5\,nm. A comparison of the thermal conductance of AGHs and GaAs bulk is shown in Fig.\,\ref{Fig3}. It establishes firstly that the thermal conductance of the AGHs is orders of magnitude smaller and secondly that the slope of the temperature dependence is inversed. From these results, we conclude that the thermal conductance of the AGHs is determined by different mechanisms compared to the  bulk  material

To understand the measured thermal conductances of the AGHs we assume a simple model, in which the thermal conductance is mainly determined by phonon transport through the pillars. At $T$=300\,K  the mfp of the phonons in GaAs is 145.3\,nm \cite{Che98}. Since the pillars are much shorter than the mfp, we assume that the phonon current through the pillars is not influenced by phonon scattering but instead it is determined by the probability that a phonon passes through the pillars. This scenario is analogous to a current of gas molecules that passes through a hole in a containment or to ballistic electrons passing a Sharvin-point contact\,\cite{Sha64}. Thus, in our model each pillar represents a thermal point contact that connects two half spaces with a phonon gas. In each half space, the phonon gas is in equilibrium at a certain temperature. This assumption is reasonable, because the thermal conductances of the capping layer and the substrate are orders of magnitude larger compared to the air gap. The heat current $I_{th}$ through a single pillar is given by the Boltzmann transport equation \cite{Ang98, Pra06, Pra07, Che98, Zhou09}
\begin{equation}
 \begin{split}
 I_{th}= A\sum_{p=1}^{3} \int_{k=0}^{k_{max}} \frac{1}{(2\pi)^3}[f(k, T_{h})-f(k, T_{c})]\\
 \cdot \hbar \omega_{p}(k) v_p(k) \cos(\theta)d^3k,
  \end{split}
  \label{Eq3}
  \end{equation}
where $A$ is the cross sectional area of the pillar, $k$ is the wave number of the phonons, $f(k, T)$ is the Bose-Einstein distribution at temperatures $T_{h}$ and $T_{c}$ of the hot and cold half space, respectively.  The phonon energy is $\hbar \omega_{p}$ and $v_p\cos(\theta)$ is the phonon group velocity component in the direction of the heat current. The sum 
considers  three phonon polarizations $p$, one longitudinal and two transversal acoustic phonon modes. The optical phonon modes are neglected, because of their low velocities and high activation energies\,\cite{Che98}. To simplify the integral in eq.\,(\ref{Eq3}), we approximate the dispersion relation for the fcc-GaAs crystal in (001)\,direction by 
\begin{eqnarray}
\omega_{p}(k)=\omega_{p}^{max} \sin\left(\frac{ka}{4}\right),
\label{Eq4}
\end{eqnarray}
with the lattice constant $a$ and the maximum phonon frequency $\omega_{p}^{max}$. The values of $\omega_{p}^{max}$ are taken from Ref.\,\cite{Bla82}. Using eq.\,(\ref{Eq2}) and  substituting $x_p=\hbar \omega_{p}/k_BT$ we finally get
\begin{multline}
K_{pillar}=\frac{I_{th}}{\Delta T}= \\²
\frac{2}{\pi^2}\frac{k_B^2T}{\hbar}\frac{A}{a^2}\sum_{p=1}^{3}\int_{x_p=0}^{x_p^{max}} [\arcsin(x_p/x_p^{max})]^2  \frac{x_p^2e^{x_p}}{(e^{x_p}-1)^2} dx_p,
\label{Eq5}
\end{multline}
for the thermal conductance of a single pillar. To solve this equation the cross sectional area $A$ of the pillars is required. Moreover, since we measure the thermal conductance of an ensemble of pillars, the pillar density is needed to quantitatively compare the model results with the measurements. 

\begin{figure}[bt]
\centering
\includegraphics[width=8.5cm]{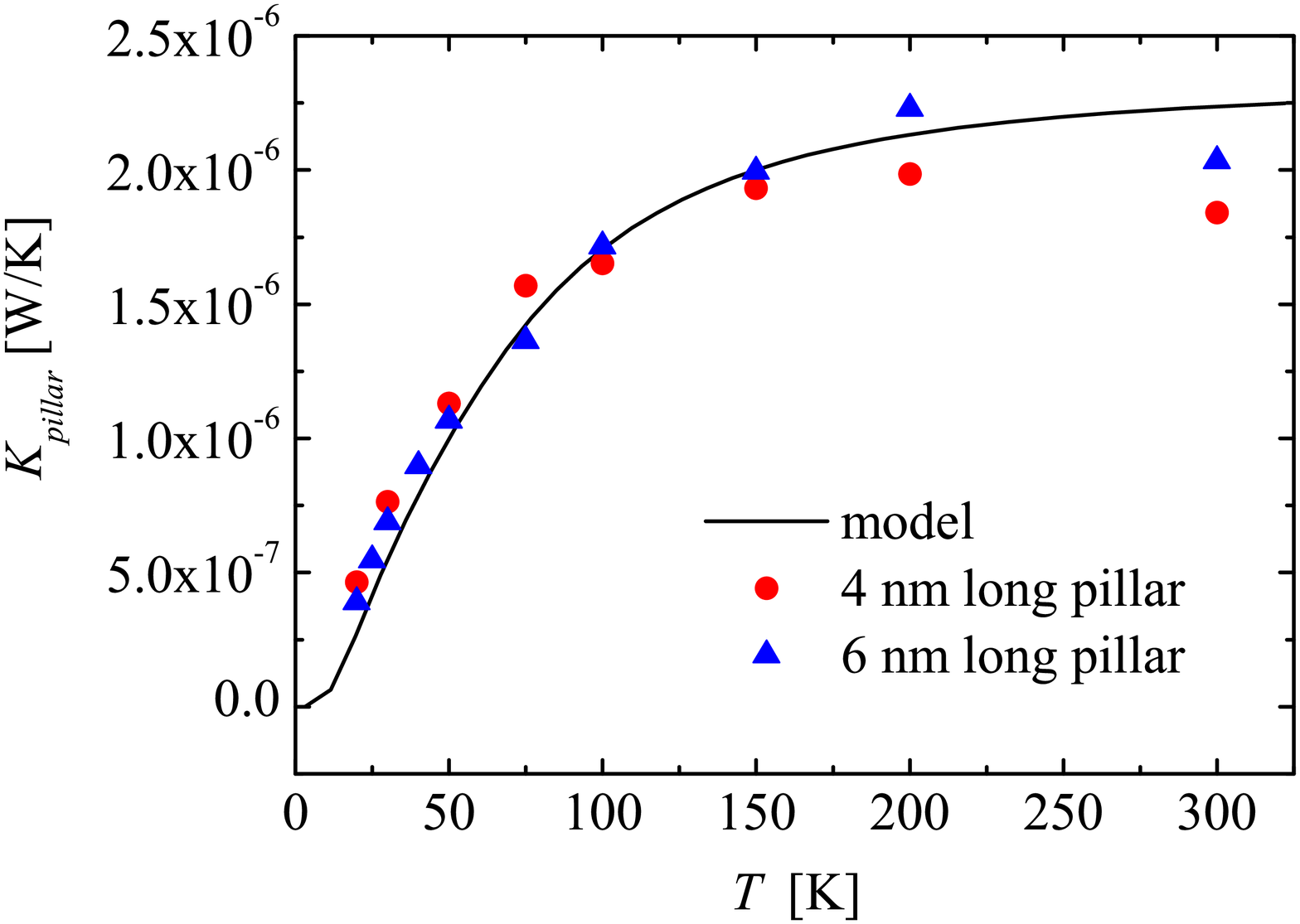}
\caption{Thermal conductance of a single pillar. The full line is calculated according to the model eq.\,(\ref{Eq5}), assuming a pillar diameter of 100\,nm.
 As described in the text, the points are calculated from the data in Fig.\,\ref{Fig3} assuming slightly different pillar densities for AGHs with 4\,nm and 6\,nm long pillars, respectively. }
\label{Fig4}
\end{figure}

Due to the fabrication method the structural parameters of the pillars are determined by the initial nanoholes. So far, measurements on AlAs surfaces are not available, which is due to the very fast oxidation of the highly reactive AlAs under air \cite{Hey094}. Therefore, we estimate the structural parameters of the pillars from atomic force microscopy (AFM) of nanoholes created by LDE in Al$_{0.35}$Ga$_{0.65}$As and GaAs surfaces. In previous publications the total hole density was of about 2\,$\mu$m$^{-2}$ \cite{Hey09} and the average diameter was 100\,nm \cite{Hey11, Ste08}. The AFM data show in addition a broad hole depth-distribution with a high density of holes with depth between 4\,nm and 6\,nm \cite{Hey092}.  Those holes contribute to the density of the pillars in the 4\,nm AGH sample but not to the 6\,nm AGH. This qualitatively explains the different thermal conductances measured  here for 4\,nm and 6\,nm AGH samples. While in perfect ballistic point contacts a pillar length independent conduction is expected, our raw data show slightly different conductions, which we attribute to the different densities of participating pillars as describe above. Due to the uncertainties of the hole densities and depth distribution on AlAs, we consider the respective pillars densities in 4\,nm and 6\,nm AGH as fitting parameter, while we take the diameter fix on 100\,nm, to compare the measurements with the model. We obtain best agreement using a density of 6.4\,$\mu$m$^{-2}$ for the 4\,nm long and of 3.75\,$\mu$m$^{-2}$ for the 6\,nm long pillars. The results are shown in Fig.\,\ref{Fig4}.
The higher pillar densities determined here is probably related to a reduced coefficient for surface diffusion of Ga droplets on AlAs surfaces in comparison to Al$_{0.35}$Ga$_{0.65}$As and GaAs surfaces  and a thus increased hole density. The very good agreement between model results and experimental data indicates that the  thermal conductance of the AGHs is dominated by the proposed mechanism of ballistic heat transport through the pillars.

 The key result that the insertion of an air gap with low density nanopillars reduces thermal conductance by up to five orders of magnitude, may pave the way to novel thermoelectric devices with large efficiencies \cite{Hoc07,Har02, Hsu04, Pou08, Ven01, Vin10, Mar04, Zeng06, Wes08}. Because of the adjustable small gap sizes and the epitaxial relationship between capping layer and substrate the AGHs are probably useful for investigations of thermionic effects\,\cite{Zeng06, Wes08}. Interestingly, the thermal conductance through a single pillar is even up to three orders of magnitude reduced in comparison to a structure with the pillar size but with the specific thermal conductivity of GaAs bulk material. Thus the pillars by their own are interesting for thermoelectric investigations \cite{Hoc07}.

In conclusion, we have measured the thermal conductance of short GaAs nanopillars. The lengths of the pillars are well below the phonon mfp up to room temperature. A simple model of ballistic thermal transport through point contacts was introduced to explain the experimental results. The model exhibits very good qualitative agreement with the experimental conductance data.  The agreement is also quantitatively good, if we assume pillar diameters and densities that are in the range of earlier measurements. This supports our approach of ballistic thermal transport in the pillars. Furthermore, the observed thermal conductances are orders of magnitude smaller than in comparable structures with specific thermal conductivity of GaAs bulk material. Based on this reduction, the AGHs may open the doorway for new thermoelectric devices with enhanced efficiencies \cite{Hoc07,Har02, Hsu04, Pou08, Ven01, Vin10, Mar04, Zeng06, Wes08}.  Electrical transport studies through doped pillars are pending for thermo-power studies.
 
This work was supported by the Deutsche Forschungsgemeinschaft via HA 2042/6-1, GrK 1286 
and SSP 1386. 
Also we thank David Sonnenberg for helpful discussions and the growth of samples for the determination of the pillar structural parameters.

\begin{thebibliography}{28}%
\makeatletter
\providecommand \@ifxundefined [1]{%
 \@ifx{#1\undefined}
}%
\providecommand \@ifnum [1]{%
 \ifnum #1\expandafter \@firstoftwo
 \else \expandafter \@secondoftwo
 \fi
}%
\providecommand \@ifx [1]{%
 \ifx #1\expandafter \@firstoftwo
 \else \expandafter \@secondoftwo
 \fi
}%
\providecommand \natexlab [1]{#1}%
\providecommand \enquote  [1]{``#1''}%
\providecommand \bibnamefont  [1]{#1}%
\providecommand \bibfnamefont [1]{#1}%
\providecommand \citenamefont [1]{#1}%
\providecommand \href@noop [0]{\@secondoftwo}%
\providecommand \href [0]{\begingroup \@sanitize@url \@href}%
\providecommand \@href[1]{\@@startlink{#1}\@@href}%
\providecommand \@@href[1]{\endgroup#1\@@endlink}%
\providecommand \@sanitize@url [0]{\catcode `\\12\catcode `\$12\catcode
  `\&12\catcode `\#12\catcode `\^12\catcode `\_12\catcode `\%12\relax}%
\providecommand \@@startlink[1]{}%
\providecommand \@@endlink[0]{}%
\providecommand \url  [0]{\begingroup\@sanitize@url \@url }%
\providecommand \@url [1]{\endgroup\@href {#1}{\urlprefix }}%
\providecommand \urlprefix  [0]{URL }%
\providecommand \Eprint [0]{\href }%
\providecommand \doibase [0]{http://dx.doi.org/}%
\providecommand \selectlanguage [0]{\@gobble}%
\providecommand \bibinfo  [0]{\@secondoftwo}%
\providecommand \bibfield  [0]{\@secondoftwo}%
\providecommand \translation [1]{[#1]}%
\providecommand \BibitemOpen [0]{}%
\providecommand \bibitemStop [0]{}%
\providecommand \bibitemNoStop [0]{.\EOS\space}%
\providecommand \EOS [0]{\spacefactor3000\relax}%
\providecommand \BibitemShut  [1]{\csname bibitem#1\endcsname}%
\let\auto@bib@innerbib\@empty
\bibitem [{\citenamefont {Schwab}\ \emph {et~al.}(2000)\citenamefont {Schwab},
  \citenamefont {Henriksen}, \citenamefont {Worlock},\ and\ \citenamefont
  {Roukes}}]{Sch00}%
  \BibitemOpen
  \bibfield  {author} {\bibinfo {author} {\bibfnamefont {K.}~\bibnamefont
  {Schwab}}, \bibinfo {author} {\bibfnamefont {E.~A.}\ \bibnamefont
  {Henriksen}}, \bibinfo {author} {\bibfnamefont {J.~M.}\ \bibnamefont
  {Worlock}}, \ and\ \bibinfo {author} {\bibfnamefont {M.~L.}\ \bibnamefont
  {Roukes}},\ }\href@noop {} {\bibfield  {journal} {\bibinfo  {journal}
  {Nature}\ }\textbf {\bibinfo {volume} {404}},\ \bibinfo {pages} {974}
  (\bibinfo {year} {2000})}\BibitemShut {NoStop}%
\bibitem [{\citenamefont {Hochbaum}\ \emph {et~al.}(2007)\citenamefont
  {Hochbaum}, \citenamefont {Chen}, \citenamefont {Delgardo}, \citenamefont
  {Liang}, \citenamefont {Garnett}, \citenamefont {Najarian}, \citenamefont
  {Majumdar},\ and\ \citenamefont {Yang}}]{Hoc07}%
  \BibitemOpen
  \bibfield  {author} {\bibinfo {author} {\bibfnamefont {A.~I.}\ \bibnamefont
  {Hochbaum}}, \bibinfo {author} {\bibfnamefont {R.}~\bibnamefont {Chen}},
  \bibinfo {author} {\bibfnamefont {R.}~\bibnamefont {Delgardo}}, \bibinfo
  {author} {\bibfnamefont {W.}~\bibnamefont {Liang}}, \bibinfo {author}
  {\bibfnamefont {E.}~\bibnamefont {Garnett}}, \bibinfo {author} {\bibfnamefont
  {M.}~\bibnamefont {Najarian}}, \bibinfo {author} {\bibfnamefont
  {A.}~\bibnamefont {Majumdar}}, \ and\ \bibinfo {author} {\bibfnamefont
  {P.}~\bibnamefont {Yang}},\ }\href@noop {} {\bibfield  {journal} {\bibinfo
  {journal} {Nature Lett.}\ }\textbf {\bibinfo {volume} {451}},\ \bibinfo
  {pages} {163} (\bibinfo {year} {2007})}\BibitemShut {NoStop}%
\bibitem [{\citenamefont {Harman}\ \emph {et~al.}(2002)\citenamefont {Harman},
  \citenamefont {Taylor}, \citenamefont {Walsh},\ and\ \citenamefont
  {Laforge}}]{Har02}%
  \BibitemOpen
  \bibfield  {author} {\bibinfo {author} {\bibfnamefont {T.}~\bibnamefont
  {Harman}}, \bibinfo {author} {\bibfnamefont {P.}~\bibnamefont {Taylor}},
  \bibinfo {author} {\bibfnamefont {M.}~\bibnamefont {Walsh}}, \ and\ \bibinfo
  {author} {\bibfnamefont {B.}~\bibnamefont {Laforge}},\ }\href@noop {}
  {\bibfield  {journal} {\bibinfo  {journal} {Science}\ }\textbf {\bibinfo
  {volume} {297}},\ \bibinfo {pages} {2229} (\bibinfo {year}
  {2002})}\BibitemShut {NoStop}%
\bibitem [{\citenamefont {Hsu}\ \emph {et~al.}(2004)\citenamefont {Hsu},
  \citenamefont {Loo}, \citenamefont {Guo}, \citenamefont {Chen}, \citenamefont
  {Dyck}, \citenamefont {Uher}, \citenamefont {Hogan}, \citenamefont
  {Polychroniadis},\ and\ \citenamefont {Kanatzidis}}]{Hsu04}%
  \BibitemOpen
  \bibfield  {author} {\bibinfo {author} {\bibfnamefont {K.~F.}\ \bibnamefont
  {Hsu}}, \bibinfo {author} {\bibfnamefont {S.}~\bibnamefont {Loo}}, \bibinfo
  {author} {\bibfnamefont {F.}~\bibnamefont {Guo}}, \bibinfo {author}
  {\bibfnamefont {W.}~\bibnamefont {Chen}}, \bibinfo {author} {\bibfnamefont
  {J.~S.}\ \bibnamefont {Dyck}}, \bibinfo {author} {\bibfnamefont
  {C.}~\bibnamefont {Uher}}, \bibinfo {author} {\bibfnamefont {T.}~\bibnamefont
  {Hogan}}, \bibinfo {author} {\bibfnamefont {E.~K.}\ \bibnamefont
  {Polychroniadis}}, \ and\ \bibinfo {author} {\bibfnamefont {M.~G.}\
  \bibnamefont {Kanatzidis}},\ }\href {\doibase 10.1126/science.1092963} {\
  \textbf {\bibinfo {volume} {303}},\ \bibinfo {pages} {818} (\bibinfo {year}
  {2004})}\BibitemShut {NoStop}%
\bibitem [{\citenamefont {Poudel}\ \emph {et~al.}(2008)\citenamefont {Poudel},
  \citenamefont {Hao}, \citenamefont {Ma}, \citenamefont {Lan}, \citenamefont
  {Minnich}, \citenamefont {Yu}, \citenamefont {Yan}, \citenamefont {Wang},
  \citenamefont {Muto}, \citenamefont {Vashaee}, \citenamefont {Chen},
  \citenamefont {Liu}, \citenamefont {Dresselhaus}, \citenamefont {Chen},\ and\
  \citenamefont {Ren}}]{Pou08}%
  \BibitemOpen
  \bibfield  {author} {\bibinfo {author} {\bibfnamefont {B.}~\bibnamefont
  {Poudel}}, \bibinfo {author} {\bibfnamefont {Q.}~\bibnamefont {Hao}},
  \bibinfo {author} {\bibfnamefont {Y.}~\bibnamefont {Ma}}, \bibinfo {author}
  {\bibfnamefont {Y.}~\bibnamefont {Lan}}, \bibinfo {author} {\bibfnamefont
  {A.}~\bibnamefont {Minnich}}, \bibinfo {author} {\bibfnamefont
  {B.}~\bibnamefont {Yu}}, \bibinfo {author} {\bibfnamefont {X.}~\bibnamefont
  {Yan}}, \bibinfo {author} {\bibfnamefont {D.}~\bibnamefont {Wang}}, \bibinfo
  {author} {\bibfnamefont {A.}~\bibnamefont {Muto}}, \bibinfo {author}
  {\bibfnamefont {D.}~\bibnamefont {Vashaee}}, \bibinfo {author} {\bibfnamefont
  {X.}~\bibnamefont {Chen}}, \bibinfo {author} {\bibfnamefont {J.}~\bibnamefont
  {Liu}}, \bibinfo {author} {\bibfnamefont {M.~S.}\ \bibnamefont
  {Dresselhaus}}, \bibinfo {author} {\bibfnamefont {G.}~\bibnamefont {Chen}}, \
  and\ \bibinfo {author} {\bibfnamefont {Z.}~\bibnamefont {Ren}},\ }\href
  {\doibase 10.1126/science.1156446} {\ \textbf {\bibinfo {volume} {320}},\
  \bibinfo {pages} {634} (\bibinfo {year} {2008})}\BibitemShut {NoStop}%
\bibitem [{\citenamefont {Venkatasubramanian}\ \emph
  {et~al.}(2001)\citenamefont {Venkatasubramanian}, \citenamefont {Siivola},
  \citenamefont {Colpitts},\ and\ \citenamefont {O´Quinn}}]{Ven01}%
  \BibitemOpen
  \bibfield  {author} {\bibinfo {author} {\bibfnamefont {R.}~\bibnamefont
  {Venkatasubramanian}}, \bibinfo {author} {\bibfnamefont {E.}~\bibnamefont
  {Siivola}}, \bibinfo {author} {\bibfnamefont {T.}~\bibnamefont {Colpitts}}, \
  and\ \bibinfo {author} {\bibfnamefont {B.}~\bibnamefont {O´Quinn}},\
  }\href@noop {} {\bibfield  {journal} {\bibinfo  {journal} {Science}\ }\textbf
  {\bibinfo {volume} {303}},\ \bibinfo {pages} {777} (\bibinfo {year}
  {2001})}\BibitemShut {NoStop}%
\bibitem [{\citenamefont {Vineis}\ \emph {et~al.}(e in)\citenamefont {Vineis},
  \citenamefont {Shakouri}, \citenamefont {Majumdar},\ and\ \citenamefont
  {Kanatzidis}}]{Vin10}%
  \BibitemOpen
  \bibfield  {author} {\bibinfo {author} {\bibfnamefont {C.~J.}\ \bibnamefont
  {Vineis}}, \bibinfo {author} {\bibfnamefont {A.}~\bibnamefont {Shakouri}},
  \bibinfo {author} {\bibfnamefont {A.}~\bibnamefont {Majumdar}}, \ and\
  \bibinfo {author} {\bibfnamefont {M.~G.}\ \bibnamefont {Kanatzidis}},\ }\href
  {\doibase 10.1002/adma.201000839} {\bibfield  {journal} {\bibinfo  {journal}
  {Adv. Mat.}\ }\textbf {\bibinfo {volume} {22}},\ \bibinfo {pages} {3970}
  (\bibinfo {year} {2010)(and the references there in})}\BibitemShut {NoStop}%
\bibitem [{\citenamefont {Majumdar}(2004)}]{Mar04}%
  \BibitemOpen
  \bibfield  {author} {\bibinfo {author} {\bibfnamefont {A.}~\bibnamefont
  {Majumdar}},\ }\href@noop {} {\bibfield  {journal} {\bibinfo  {journal}
  {Science}\ }\textbf {\bibinfo {volume} {303}},\ \bibinfo {pages} {777}
  (\bibinfo {year} {2004})}\BibitemShut {NoStop}%
\bibitem [{\citenamefont {Zeng}(2006)}]{Zeng06}%
  \BibitemOpen
  \bibfield  {author} {\bibinfo {author} {\bibfnamefont {T.}~\bibnamefont
  {Zeng}},\ }\href@noop {} {\bibfield  {journal} {\bibinfo  {journal} {Appl.
  Phys. Lett.}\ }\textbf {\bibinfo {volume} {88}},\ \bibinfo {pages} {153104}
  (\bibinfo {year} {2006})}\BibitemShut {NoStop}%
\bibitem [{\citenamefont {Westover}\ and\ \citenamefont {Fisher}(e
  in)}]{Wes08}%
  \BibitemOpen
  \bibfield  {author} {\bibinfo {author} {\bibfnamefont {T.~L.}\ \bibnamefont
  {Westover}}\ and\ \bibinfo {author} {\bibfnamefont {T.~S.}\ \bibnamefont
  {Fisher}},\ }\href {\doibase 10.1103/PhysRevB.77.115426} {\bibfield
  {journal} {\bibinfo  {journal} {Phys. Rev. B}\ }\textbf {\bibinfo {volume}
  {77}},\ \bibinfo {pages} {115426} (\bibinfo {year} {2008) (and the references
  there in})}\BibitemShut {NoStop}%
\bibitem [{\citenamefont {Angelescu}\ \emph {et~al.}(1998)\citenamefont
  {Angelescu}, \citenamefont {Cross},\ and\ \citenamefont {Roukes}}]{Ang98}%
  \BibitemOpen
  \bibfield  {author} {\bibinfo {author} {\bibfnamefont {D.~E.}\ \bibnamefont
  {Angelescu}}, \bibinfo {author} {\bibfnamefont {M.~C.}\ \bibnamefont
  {Cross}}, \ and\ \bibinfo {author} {\bibfnamefont {M.~L.}\ \bibnamefont
  {Roukes}},\ }\href {\doibase DOI: 10.1006/spmi.1997.0561} {\bibfield
  {journal} {\bibinfo  {journal} {Superlattices and Microstructures}\ }\textbf
  {\bibinfo {volume} {23}},\ \bibinfo {pages} {673 } (\bibinfo {year}
  {1998})}\BibitemShut {NoStop}%
\bibitem [{\citenamefont {Chen}(e in)}]{Che98}%
  \BibitemOpen
  \bibfield  {author} {\bibinfo {author} {\bibfnamefont {G.}~\bibnamefont
  {Chen}},\ }\href@noop {} {\bibfield  {journal} {\bibinfo  {journal} {Phys.
  Rev. B}\ }\textbf {\bibinfo {volume} {57}},\ \bibinfo {pages} {14958}
  (\bibinfo {year} {1998) (and the references there in})}\BibitemShut {NoStop}%
\bibitem [{\citenamefont {Prasher}(2006)}]{Pra06}%
  \BibitemOpen
  \bibfield  {author} {\bibinfo {author} {\bibfnamefont {R.}~\bibnamefont
  {Prasher}},\ }\href@noop {} {\bibfield  {journal} {\bibinfo  {journal} {Phys.
  Rev. B}\ }\textbf {\bibinfo {volume} {74}},\ \bibinfo {pages} {165413}
  (\bibinfo {year} {2006})}\BibitemShut {NoStop}%
\bibitem [{\citenamefont {Prasher}\ \emph {et~al.}(2008)\citenamefont
  {Prasher}, \citenamefont {Tong},\ and\ \citenamefont {Majumdar}}]{Pra07}%
  \BibitemOpen
  \bibfield  {author} {\bibinfo {author} {\bibfnamefont {R.}~\bibnamefont
  {Prasher}}, \bibinfo {author} {\bibfnamefont {T.}~\bibnamefont {Tong}}, \
  and\ \bibinfo {author} {\bibfnamefont {A.}~\bibnamefont {Majumdar}},\ }\href
  {\doibase 10.1021/nl0721665} {\bibfield  {journal} {\bibinfo  {journal} {Nano
  Letters}\ }\textbf {\bibinfo {volume} {8}},\ \bibinfo {pages} {99} (\bibinfo
  {year} {2008})}\BibitemShut {NoStop}%
\bibitem [{\citenamefont {Zhou}\ \emph {et~al.}(2009)\citenamefont {Zhou},
  \citenamefont {Wang}, \citenamefont {Zhu}, \citenamefont {Peng},\ and\
  \citenamefont {Chen}}]{Zhou09}%
  \BibitemOpen
  \bibfield  {author} {\bibinfo {author} {\bibfnamefont {L.-P.}\ \bibnamefont
  {Zhou}}, \bibinfo {author} {\bibfnamefont {M.-P.}\ \bibnamefont {Wang}},
  \bibinfo {author} {\bibfnamefont {J.-J.}\ \bibnamefont {Zhu}}, \bibinfo
  {author} {\bibfnamefont {X.-F.}\ \bibnamefont {Peng}}, \ and\ \bibinfo
  {author} {\bibfnamefont {K.-Q.}\ \bibnamefont {Chen}},\ }\href {\doibase
  DOI:10.1063/1.3142302} {\bibfield  {journal} {\bibinfo  {journal} {J. Appl.
  Phys.}\ }\textbf {\bibinfo {volume} {105}},\ \bibinfo {pages} {114318}
  (\bibinfo {year} {2009})}\BibitemShut {NoStop}%
\bibitem [{\citenamefont {Heyn}\ \emph {et~al.}(2011)\citenamefont {Heyn},
  \citenamefont {Schmidt}, \citenamefont {Schwaiger}, \citenamefont {Stemmann},
  \citenamefont {Mendach},\ and\ \citenamefont {Hansen}}]{Hey10}%
  \BibitemOpen
  \bibfield  {author} {\bibinfo {author} {\bibfnamefont {C.}~\bibnamefont
  {Heyn}}, \bibinfo {author} {\bibfnamefont {M.}~\bibnamefont {Schmidt}},
  \bibinfo {author} {\bibfnamefont {S.}~\bibnamefont {Schwaiger}}, \bibinfo
  {author} {\bibfnamefont {A.}~\bibnamefont {Stemmann}}, \bibinfo {author}
  {\bibfnamefont {S.}~\bibnamefont {Mendach}}, \ and\ \bibinfo {author}
  {\bibfnamefont {W.}~\bibnamefont {Hansen}},\ }\href {\doibase
  10.1063/1.3544047} {\bibfield  {journal} {\bibinfo  {journal} {Appl. Phys.
  Lett.}\ }\textbf {\bibinfo {volume} {98}},\ \bibinfo {eid} {033105} (\bibinfo
  {year} {2011})}\BibitemShut {NoStop}%
\bibitem [{\citenamefont {Heyn}\ \emph
  {et~al.}(2009{\natexlab{a}})\citenamefont {Heyn}, \citenamefont {Stemmann},
  \citenamefont {K\"{o}ppen}, \citenamefont {Strelow}, \citenamefont {Kipp},
  \citenamefont {Grave}, \citenamefont {Mendach},\ and\ \citenamefont
  {Hansen}}]{Hey094}%
  \BibitemOpen
  \bibfield  {author} {\bibinfo {author} {\bibfnamefont {C.}~\bibnamefont
  {Heyn}}, \bibinfo {author} {\bibfnamefont {A.}~\bibnamefont {Stemmann}},
  \bibinfo {author} {\bibfnamefont {T.}~\bibnamefont {K\"{o}ppen}}, \bibinfo
  {author} {\bibfnamefont {C.}~\bibnamefont {Strelow}}, \bibinfo {author}
  {\bibfnamefont {T.}~\bibnamefont {Kipp}}, \bibinfo {author} {\bibfnamefont
  {M.}~\bibnamefont {Grave}}, \bibinfo {author} {\bibfnamefont
  {S.}~\bibnamefont {Mendach}}, \ and\ \bibinfo {author} {\bibfnamefont
  {W.}~\bibnamefont {Hansen}},\ }\href {\doibase 10.1063/1.3133338} {\bibfield
  {journal} {\bibinfo  {journal} {Appl. Phys. Lett.}\ }\textbf {\bibinfo
  {volume} {94}},\ \bibinfo {eid} {183113} (\bibinfo {year}
  {2009}{\natexlab{a}})}\BibitemShut {NoStop}%
\bibitem [{\citenamefont {Heyn}\ \emph
  {et~al.}(2009{\natexlab{b}})\citenamefont {Heyn}, \citenamefont {Stemmann},\
  and\ \citenamefont {Hansen}}]{Hey09}%
  \BibitemOpen
  \bibfield  {author} {\bibinfo {author} {\bibfnamefont {C.}~\bibnamefont
  {Heyn}}, \bibinfo {author} {\bibfnamefont {A.}~\bibnamefont {Stemmann}}, \
  and\ \bibinfo {author} {\bibfnamefont {W.}~\bibnamefont {Hansen}},\ }\href
  {\doibase 10.1063/1.3254216} {\bibfield  {journal} {\bibinfo  {journal}
  {Appl. Phys. Lett.}\ }\textbf {\bibinfo {volume} {95}},\ \bibinfo {eid}
  {173110} (\bibinfo {year} {2009}{\natexlab{b}})}\BibitemShut {NoStop}%
\bibitem [{\citenamefont {Heyn}(2011)}]{Hey11}%
  \BibitemOpen
  \bibfield  {author} {\bibinfo {author} {\bibfnamefont {C.}~\bibnamefont
  {Heyn}},\ }\href {\doibase 10.1103/PhysRevB.83.165302} {\bibfield  {journal}
  {\bibinfo  {journal} {Phys. Rev. B}\ }\textbf {\bibinfo {volume} {83}},\
  \bibinfo {pages} {165302} (\bibinfo {year} {2011})}\BibitemShut {NoStop}%
\bibitem [{\citenamefont {Stemmann}\ \emph {et~al.}(2008)\citenamefont
  {Stemmann}, \citenamefont {Heyn}, \citenamefont {K\"{o}ppen}, \citenamefont
  {Kipp},\ and\ \citenamefont {Hansen}}]{Ste08}%
  \BibitemOpen
  \bibfield  {author} {\bibinfo {author} {\bibfnamefont {A.}~\bibnamefont
  {Stemmann}}, \bibinfo {author} {\bibfnamefont {C.}~\bibnamefont {Heyn}},
  \bibinfo {author} {\bibfnamefont {T.}~\bibnamefont {K\"{o}ppen}}, \bibinfo
  {author} {\bibfnamefont {T.}~\bibnamefont {Kipp}}, \ and\ \bibinfo {author}
  {\bibfnamefont {W.}~\bibnamefont {Hansen}},\ }\href {\doibase
  10.1063/1.2981517} {\bibfield  {journal} {\bibinfo  {journal} {Appl. Phys.
  Lett.}\ }\textbf {\bibinfo {volume} {93}},\ \bibinfo {eid} {123108} (\bibinfo
  {year} {2008})}\BibitemShut {NoStop}%
\bibitem [{\citenamefont {Heyn}\ \emph
  {et~al.}(2009{\natexlab{c}})\citenamefont {Heyn}, \citenamefont {Stemmann},\
  and\ \citenamefont {Hansen}}]{Hey092}%
  \BibitemOpen
  \bibfield  {author} {\bibinfo {author} {\bibfnamefont {C.}~\bibnamefont
  {Heyn}}, \bibinfo {author} {\bibfnamefont {A.}~\bibnamefont {Stemmann}}, \
  and\ \bibinfo {author} {\bibfnamefont {W.}~\bibnamefont {Hansen}},\
  }\href@noop {} {\bibfield  {journal} {\bibinfo  {journal} {J. of Crys.
  Growth}\ }\textbf {\bibinfo {volume} {311}},\ \bibinfo {pages} {1839 }
  (\bibinfo {year} {2009}{\natexlab{c}})}\BibitemShut {NoStop}%
\bibitem [{\citenamefont {Cahill}(1990)}]{Cah89}%
  \BibitemOpen
  \bibfield  {author} {\bibinfo {author} {\bibfnamefont {D.}~\bibnamefont
  {Cahill}},\ }\href@noop {} {\bibfield  {journal} {\bibinfo  {journal} {Rev.
  Sci. Instrum.}\ }\textbf {\bibinfo {volume} {61}},\ \bibinfo {pages} {802}
  (\bibinfo {year} {1990})}\BibitemShut {NoStop}%
\bibitem [{\citenamefont {Cahill}\ \emph {et~al.}(1994)\citenamefont {Cahill},
  \citenamefont {Katiyar},\ and\ \citenamefont {Abelson.}}]{Cah94}%
  \BibitemOpen
  \bibfield  {author} {\bibinfo {author} {\bibfnamefont {D.~G.}~\bibnamefont
  {Cahill}}, \bibinfo {author} {\bibfnamefont {M.}~\bibnamefont {Katiyar}}, \
  and\ \bibinfo {author} {\bibfnamefont {J.~R.}~\ \bibnamefont {Abelson}},\
  }\href@noop {} {\bibfield  {journal} {\bibinfo  {journal} {Phys. Rev. B}\
  }\textbf {\bibinfo {volume} {50}},\ \bibinfo {pages} {6077} (\bibinfo {year}
  {1994})}\BibitemShut {NoStop}%
\bibitem [{\citenamefont {Borca-Tasciuc}\ \emph {et~al.}(2001)\citenamefont
  {Borca-Tasciuc}, \citenamefont {Kumar},\ and\ \citenamefont {Chen}}]{Bor01}%
  \BibitemOpen
  \bibfield  {author} {\bibinfo {author} {\bibfnamefont {T.}~\bibnamefont
  {Borca-Tasciuc}}, \bibinfo {author} {\bibfnamefont {A.~R.}\ \bibnamefont
  {Kumar}}, \ and\ \bibinfo {author} {\bibfnamefont {G.}~\bibnamefont {Chen}},\
  }\href@noop {} {\bibfield  {journal} {\bibinfo  {journal} {Rev. Sci.
  Instrum.}\ }\textbf {\bibinfo {volume} {72}},\ \bibinfo {pages} {2139}
  (\bibinfo {year} {2001})}\BibitemShut {NoStop}%
\bibitem [{\citenamefont {Carlson}\ \emph {et~al.}(1965)\citenamefont
  {Carlson}, \citenamefont {Slack},\ and\ \citenamefont {Silverman}}]{Car65}%
  \BibitemOpen
  \bibfield  {author} {\bibinfo {author} {\bibfnamefont {R.~O.}\ \bibnamefont
  {Carlson}}, \bibinfo {author} {\bibfnamefont {G.}~\bibnamefont {Slack}}, \
  and\ \bibinfo {author} {\bibfnamefont {S.}~\bibnamefont {Silverman}},\
  }\href@noop {} {\bibfield  {journal} {\bibinfo  {journal} {J. Appl. Phys.}\
  }\textbf {\bibinfo {volume} {36}},\ \bibinfo {pages} {505} (\bibinfo {year}
  {1965})}\BibitemShut {NoStop}%
\bibitem [{\citenamefont {Maycock}(1967)}]{May66}%
  \BibitemOpen
  \bibfield  {author} {\bibinfo {author} {\bibfnamefont {P.~D.}\ \bibnamefont
  {Maycock}},\ }\href@noop {} {\bibfield  {journal} {\bibinfo  {journal}
  {Sol.-Sta. Elec. Perg. Press}\ }\textbf {\bibinfo {volume} {10}},\ \bibinfo
  {pages} {161} (\bibinfo {year} {1967})}\BibitemShut {NoStop}%
\bibitem [{\citenamefont {Sharvin}(1965)}]{Sha64}%
  \BibitemOpen
  \bibfield  {author} {\bibinfo {author} {\bibfnamefont {Y.~V.}\ \bibnamefont
  {Sharvin}},\ }\href@noop {} {\bibfield  {journal} {\bibinfo  {journal} {J.
  Exptl. Theoret. Phys.}\ }\textbf {\bibinfo {volume} {48}},\ \bibinfo {pages}
  {984} (\bibinfo {year} {1965})}\BibitemShut {NoStop}%
\bibitem [{\citenamefont {Blackemore}(1982)}]{Bla82}%
  \BibitemOpen
  \bibfield  {author} {\bibinfo {author} {\bibfnamefont {J.~S.}\ \bibnamefont
  {Blackemore}},\ }\href@noop {} {\bibfield  {journal} {\bibinfo  {journal} {J.
  Appl. Phys.}\ }\textbf {\bibinfo {volume} {53}},\ \bibinfo {pages} {R123}
  (\bibinfo {year} {1982})}\BibitemShut {NoStop}%
\end{thebibliography}
%

\end{document}